\documentstyle[12pt]{article}

\makeatletter
\def\vereq#1#2{\lower3pt\vbox{\baselineskip1.5pt \lineskip1.5pt
\ialign{$\m@th#1\hfill##\hfil$\crcr#2\crcr\sim\crcr}}}
\makeatother

\makeatletter

\def\vereq#1#2{\lower3pt\vbox{\baselineskip1.5pt \lineskip1.5pt
\ialign{$\m@th#1\hfill##\hfil$\crcr#2\crcr\sim\crcr}}}

\makeatother

\newcommand{\beq}{\begin{equation}}
\newcommand{\eeq}{\end{equation}}

\newcommand{\remove}[1]{}

\begin{document}
\begin{titlepage}
\begin{center}
\today     \hfill    LBL-38047\\
{}~{} \hfill UCB-PTH-95/43\\

\vskip .25in

{\large \bf $(S_3)^3$ flavor symmetry and $p\rightarrow K^0 e^+$}
\footnote{This work was supported in part by the Director, Office of
Energy Research, Office of High Energy and Nuclear Physics, Division of
High Energy Physics of the U.S. Department of Energy under Contract
DE-AC03-76SF00098 and in part by the National Science Foundation under
grant PHY-90-21139.}
%
\vskip 0.3in

Christopher D. Carone,$^1$ Lawrence J. Hall,$^{1,2}$ and
Hitoshi Murayama$^{1,2}$

\vskip 0.1in

{{}$^1$ \em Theoretical Physics Group\\
     Lawrence Berkeley National Laboratory\\
     University of California, Berkeley, California 94720}

\vskip 0.1in

{{}$^2$ \em Department of Physics\\
     University of California, Berkeley, California 94720}

\end{center}

\vskip .3in

\begin{abstract}
We show how to incorporate the lepton sector in a supersymmetric
theory of flavor based on the discrete flavor group $(S_3)^3$.
Assuming that all possible nonrenormalizable operators are generated
at the Planck scale, we show that the transformation properties
of the leptons and of the flavor-symmetry breaking fields are
uniquely determined.  We then demonstrate that the model has a
viable phenomenology and makes one very striking prediction: the nucleon
decays predominantly to $K l$ where $l$ is a {\em first} generation
lepton.  We show that the modes $n \rightarrow K^0 \bar{\nu}_e$,
$p\rightarrow K^+ \bar{\nu}_e$,  and $p\rightarrow K^0 e^+$
occur at comparable rates, and could well be discovered simultaneously
at the SuperKamiokande experiment.
\end{abstract}

\end{titlepage}
\renewcommand{\thepage}{\roman{page}}
\setcounter{page}{2}
\mbox{ }

\vskip 1in

\begin{center}
{\bf Disclaimer}
\end{center}

\vskip .2in

\begin{scriptsize}
\begin{quotation}
This document was prepared as an account of work sponsored by the United
States Government. While this document is believed to contain correct
information, neither the United States Government nor any agency
thereof, nor The Regents of the University of California, nor any of their
employees, makes any warranty, express or implied, or assumes any legal
liability or responsibility for the accuracy, completeness, or usefulness
of any information, apparatus, product, or process disclosed, or represents
that its use would not infringe privately owned rights.  Reference herein
to any specific commercial products process, or service by its trade name,
trademark, manufacturer, or otherwise, does not necessarily constitute or
imply its endorsement, recommendation, or favoring by the United States
Government or any agency thereof, or The Regents of the University of
California.  The views and opinions of authors expressed herein do not
necessarily state or reflect those of the United States Government or any
agency thereof, or The Regents of the University of California.
\end{quotation}
\end{scriptsize}

\vskip 2in

\begin{center}
\begin{small}
{\it Lawrence Berkeley Laboratory is an equal opportunity employer.}
\end{small}
\end{center}

\newpage
\renewcommand{\thepage}{\arabic{page}}
\setcounter{page}{1}
\section{Introduction}

The origin of flavor has been a significant puzzle in particle physics
since the discovery of the muon.   The replication of fermion generations
and the strongly hierarchical pattern of their masses and mixing angles
is left unexplained in the Standard Model.  In most theories that
attempt to elucidate the puzzling features of flavor, new symmetries
are introduced at mass scales that are large compared to the electroweak
scale.  The new scales may be associated with the
breaking of flavor symmetries or of a grand unified gauge group.
In either case, the introduction of a very high scale in the theory
induces a large radiative correction to the Higgs mass squared, which
destabilizes the hierarchy between the high scale and the electroweak
scale.  Supersymmetry is the most promising mechanism for avoiding
this problem. Therefore, it is natural to consider the physics of flavor
in the framework of the supersymmetric standard model (SSM).

Supersymmetry, however, complicates the problem of flavor by
introducing a new sector of particles whose masses and mixing angles
must also be understood.  While no superpartner has yet been observed, the
acceptable spectrum is constrained by low-energy processes.  Most notably, a
high degree of degeneracy is required among the light generation squarks to
suppress dangerous flavor-changing effects \cite{DG}, unless there is a strong
alignment of quark and squark eigenstates \cite{NS,LNS}.
The challenge in a supersymmetric theory of flavor is to simultaneously explain
both the suppression of flavor changing effects from the scalar sector
and the hierarchical pattern of the quark Yukawa couplings. Flavor symmetries,
spontaneously broken by a hierarchy of vacuum expectation values (vevs),
provide an interesting tool for this purpose \cite{DLK,NS}. However, there
is considerable freedom in the choice of flavor group and symmetry breaking
pattern.

In the recent literature, a number of authors have tried to meet this
challenge by constructing models of flavor based on Abelian horizontal
symmetries, often motivated by superstring theory \cite{abelian}.
However, many of these models have problems with large flavor-changing effects
\cite{DPS2}. The only surviving explicit Abelian models of which we are aware
are those of reference~\cite{NS}, which, however, rely on somewhat
ad hoc sets of charge assignments to achieve alignment. Several authors have
considered non-Abelian flavor groups, leading to near degeneracy of the
lightest two generation scalars,
including $SU(2)$ \cite{DLK}, $O(2) \times U(1)$ \cite{PS}
and $U(2)$ \cite{PT}, although the latter
appears not to solve the flavor changing problem associated with the $\epsilon$
parameter of $K$ physics.  A model has also been proposed in which the three
generations of fermions are unified into an irreducible multiplet
of a non-Abelian discrete symmetry \cite{KS}. This leads to sufficient squark
degeneracy to solve the flavor changing problem, but a heavy top quark is
not guaranteed, and results only by assuming that one of three Higgs fields
remains light after flavor symmetry breaking.

Two of the authors (LJH and HM) have advocated the use of discrete,
gauged non-Abelian family symmetries to obtain the desired degree of squark
degeneracy \cite{hallmur}: global continuous symmetries are broken by quantum
gravitational effects \cite{qg}, while gauged continuous symmetries may
generate $D$-term contributions to the squark and slepton masses that are
nonuniversal \cite{Dterm}, as in the model of reference \cite{DLK}.
It was  demonstrated in Ref.~\cite{hallmur} that
the non-Abelian discrete group, $(S_3)^3$, is a promising choice for the
flavor symmetry group of the SSM.  The group $S_3$ has both a doublet
(${\bf 2}$) and a non-trivial singlet representation (${\bf 1}_A$) into which
the three generations of fermions can be embedded. In order to construct a
viable model, three separate $S_3$ factors are required, for
the left-handed doublet fields $Q$, and the right-handed
singlet fields $U^c$ and $D^c$.  The first and second generation
fields transform as doublets, which ensures the degeneracy among the
light generation squarks.  The third generation fields must then transform
as ${\bf 1}_A$s so that the theory is free of discrete gauge anomalies.
While the group $S_3$ acts identically on three objects, the representation
structure distinguishes between the generations.  Thus,
it is possible to  choose the quantum numbers for the Higgs fields so
that only the top Yukawa coupling is allowed in the symmetry limit.
The hierarchical structure of Yukawa matrices can then be understood as a
consequence of the sequential breaking of the flavor symmetry group.  The
model is appealing on more general grounds since discrete gauge symmetries
arise naturally in superstring compactifications.

The model proposed in Ref.~\cite{hallmur}, however, did not address
the problem of flavor in the lepton sector.   The lepton Yukawa matrix is
clearly hierarchical, and its eigenvalues are similar in size to those of
the down quarks.  In addition, a high degree of degeneracy between first and
second generation sleptons (or an alignment between slepton and lepton mass
eigenstates) is required to suppress dangerous lepton
flavor-violating processes.  As far as this point is concerned, it is
reasonable to expect that the flavor structure proposed in Ref.~\cite{hallmur}
for the quarks should work equally well when applied to the lepton sector.
However, there is no guarantee that the flavor symmetry will provide an
adequate suppression of the operators which mediate proton decay.
Recall that there are two possible sources of proton decay in the SSM:
$R$-parity violating dimension-four operators (like $QDL$ or $UDD$), and
nonrenormalizable, dimension-five operators (like $QQQL$ or $UUDE$)
which are likely to be generated at the Planck scale.  In the first
case, the coefficients of the $R$-parity violating operators are forced
to be extremely small by the nonobservation of proton decay. The
operators $c_1 QsL$ and $c_2 uds$, for example, are constrained such
that $c_1 c_2 < O(10^{-26})$.  In this paper we simply assume that
$R$-parity is an exact symmetry, and these operators are not
present.\footnote{Since matter parity is non-anomalous with the minimal
SSM (MSSM) particle content, it can be considered as a discrete gauge
symmetry, and is hence preserved by quantum gravity effects.  $R$-parity
is a product of matter parity and a $2\pi$ rotation in the local Lorentz
frame.} Assuming that quantum gravity effects violate global symmetries,
however, non-renormalizable operators that conserve $R$-parity, but
violate baryon and lepton number, are presumably generated at the Planck
scale.  The coefficients of the operators $(Q_1 Q_{1,2}) Q_2 L_i/M_P$
are constrained to be no larger than $O(\lambda^8)$ with $\lambda \simeq
0.22$.  In the absence of other mechanisms to eliminate these
operators,\footnote{For example, we could impose a Peccei--Quinn
symmetry \cite{SY}, discrete symmetries \cite{IR}, or
gauge U(1)$_B$ \cite{baryon}.} an adequate theory of flavor must explain
why they are sufficiently suppressed after flavor symmetry-breaking effects
are taken into account \cite{MK}.

In this article, we extend the $(S_3)^3$ model to the lepton sector.
First, we require that the leptons transform under the same $S_3$ flavor
groups as the quarks.  This is the simplest choice given that the
ordinary Higgs fields transform nontrivially under $S_3^Q$ and $S_3^U$.
We select the transformation properties of the lepton fields
under $(S_3)^3$ so that we obtain the greatest similarity between the
lepton and down-quark Yukawa matrices.  The assignment must also
forbid all dangerous dimension-five operators in the $(S_3)^3$
symmetry limit.  We will see that the transformation properties of the
lepton fields are uniquely fixed by these requirements.  We then argue
that the fundamental sources of flavor symmetry breaking are gauge singlet
fields $\phi$ that transform in the same way as the irreducible ``blocks'' of
the quark Yukawa matrices.  We will call these fields `flavons' below.
We show from a general operator analysis that models involving
flavons with simpler transformation properties can all be excluded,
if flavor physics originates at the Planck scale.  With the
flavor symmetry breaking originating only from the Yukawa
matrices, we consider the contributions to lepton flavor violation and
proton decay.  We show that the model is consistent
with the current experimental bounds.  In addition, we show that the
dominant proton decay modes in the $(S_3)^3$ model are of the form
$p \rightarrow K l$, where $l$ is a {\em first} generation lepton.  This
is never the case in either supersymmetric or non-supersymmetric grand
unified theories.  The prediction of these rather unique modes is
exciting since the total decay rate is likely to be within the reach
of the SuperKamiokande experiment.

In the next section we make the assumptions of our analysis explicit.
In section 3, we introduce an economical form for the flavons
$\phi$, which provide an adequate description of both lepton and quark mass
matrices via dimension five interactions. We use flavor-changing and baryon
number violating phenomenology to demonstrate that this choice is the simplest
possible acceptable form for the flavons in section 5. This phenomenology is
studied in much further detail in section 6, and conclusions are drawn in
section 7.

\section{The Framework}

In this paper we construct a description of flavor, for both
quarks and leptons, based on a flavor group $(S_3)^3$, spontaneously broken
by a set of flavon fields, $\phi$. We write down an effective theory beneath
the Planck scale in which the gauge symmetry is $G_{SM} = SU(3) \times SU(2)
\times U(1)$ and $R$ parity is imposed. The field content is that of the
minimal supersymmetric model, together with a set of flavon fields, which are
necessarily all neutral under the gauge group since they take vevs much larger
than the weak scale.  The theory contains the most general $F$ and $D$ terms
consistent with $G_{SM}\times (S_3)^3 \times R_P$ with all interactions scaled
by the appropriate powers of $M_{Pl}$ and all dimensionless coefficients of
order unity. The single exception to this is the absence of a Planck scale
mass for the Higgs doublets and for the flavons. In addition the theory is
taken to possess supersymmetry breaking interactions which are the most
general soft operators consistent with the symmetries of the theory. In
particular, no universality assumptions are made to relate otherwise free
parameters. At the renormalizable level, the theory is remarkably simple: the
only $F$ terms are the Yukawa coupling for the top quark, and possible
trilinears amongst the flavon fields. The only supersymmetry breaking
terms are: the three gaugino masses, a trilinear scalar interaction involving
the top squarks coupled to a Higgs doublet, scalar masses for the
flavons, the Higgs, and for the squarks and sleptons, which are diagonal in
flavor space with the structure $(m_1^2, m_1^2, m_3^2)$. Supersymmetric
non-renormalizable $F$ terms lead to the Yukawa matrices becoming functions of
$\phi / M_{Pl}$, while the non-renormalizable $F$ and $D$ terms, which contain
supersymmetry breaking spurions, similarly lead to the trilinear $A$ terms and
the scalar masses becoming functions of $\phi / M_{Pl}$.

\section{The Basic (2,2) Model}

In the $(S_3)^3$ model of Ref.~\cite{hallmur}, the quark chiral
superfields $Q$, $U$, and $D$ are assigned to ${\bf 2}+{\bf 1_A}$
representations of $S_3^Q$, $S_3^U$ and $S_3^D$, respectively. The
first two generation fields are embedded in the doublet, for the reasons
described in the Introduction.  The Higgs fields both transform
as $({\bf 1}_A, {\bf 1}_A, {\bf 1}_S)$'s, so that the top quark Yukawa
coupling is invariant under the flavor symmetry group.  The
transformation properties of the Yukawa matrices are:
\beq
Y_U \sim \left(
\begin{array}{cc|c}
\multicolumn{2}{c|}{
({\bf \tilde{2}},{\bf \tilde{2}},{\bf 1}_S) }
& ({\bf \tilde{2}},{\bf 1}_S,{\bf 1}_S) \\ \hline
\multicolumn{2}{c|}{({\bf 1}_S,{\bf \tilde{2}},{\bf 1}_S)} &
({\bf 1}_S,{\bf 1}_S,{\bf 1}_S) \end{array} \right)
\,\,\, , \,\,\,
Y_D \sim \left(\begin{array}{cc|c}
\multicolumn{2}{c|}{
({\bf \tilde{2}},{\bf 1}_A,{\bf 2})}
& ({\bf \tilde{2}},{\bf 1}_A,{\bf 1}_A) \\ \hline
\multicolumn{2}{c|}{({\bf 1}_S,{\bf 1}_A,{\bf 2})} &
({\bf 1}_S,{\bf 1}_A,{\bf 1}_A) \end{array}\right)
\label{eq:transp}
\eeq
where we use the notation ${\bf \tilde{2}} \equiv {\bf 2} \otimes {\bf
1}_A$,\footnote{${\bf \tilde{2}} = (a,b)$ is equivalent to ${\bf 2} =
(b, -a)$.}. Note that these matrices involve at most 7 irreducible multiplets
of $(S_3)^3$.  In Ref.~\cite{hallmur}, $(S_3)^3$ was broken by only four types
of flavons: $\phi({\bf\tilde{2}}, {\bf 1}_S, {\bf 1}_S)$, $\phi({\bf
\tilde{2}}, {\bf\tilde{2}}, {\bf 1}_S)$, $\phi({\bf 1}_S, {\bf 1}_A, {\bf
1}_A)$, and $\phi({\bf \tilde{2}}, {\bf 1}_A, {\bf 2})$, the minimal number
which allows us to obtain realistic masses and mixings \cite{CM3}:
\beq
Y_U = \left( \begin{array}{cc|c}
      h_u & h_c \lambda & - h_t V_{ub}\\
      0 & h_c & - h_t V_{cb} \\ \hline
      0 & 0 & h_t
            \end{array} \right) , \,\,\,\,\,\,
Y_D = \left( \begin{array}{cc|c}
      h_d & h_s \lambda & 0\\
       0 & h_s & 0 \\ \hline 0 & 0 & h_b
            \end{array} \right),
\eeq
with $\lambda \simeq 0.22$.  We will refer to this scenario as the
basic (2,2) model below, because it involves flavons which transform as
doublets under two of the $S_3$ factors simultaneously.  Note that it is
not absolutely necessary that we generate $V_{cb}$ and $V_{ub}$ by
rotations in the up sector, as we have indicated above.  We can instead
generate $V_{cb}$ and $V_{ub}$ in the down sector (by assigning appropriate
values to the (1,3) and (2,3) elements in $Y_D$) using the breaking
parameter $\phi({\bf \tilde{2}}, {\bf1}_A, {\bf 1}_A)$ rather than
$\phi({\bf \tilde{2}}, {\bf1}_S, {\bf 1}_S)$.  This choice generally gives
us much weaker phenomenological constraints, as we will see later.  In
section 5 we study whether one can construct
$\phi({\bf \tilde{2}},{\bf\tilde{2}}, {\bf 1}_S)$
and $\phi({\bf \tilde{2}}, {\bf 1}_A, {\bf 2})$ breaking parameters
from products of flavons that transform as doublets under only one $S_3$,
and find it is not possible within the framework specified in
the previous section.
\begin{table}
\begin{center}
\begin{tabular}{c|ccc}
breaking & $S_3^Q$ & $S_3^U$ & $S_3^D$\\ \hline
$h_t V_{cb}$ & $Z_2^Q$ & $S_3^U$ & $S_3^D$\\
$h_b$ & $Z_2^Q$ &
\multicolumn{2}{c}{$(S_3^U \times S_3^D) / Z_2$} \\
$h_t V_{ub}$ & nothing &
\multicolumn{2}{c}{$(S_3^U \times S_3^D) / Z_2$} \\
$h_c$, $h_c \lambda$ & nothing &
\multicolumn{2}{c}{
$(Z_2^U \times S_3^D) / Z_2$} \\
$h_s$, $h_s \lambda$ & nothing &
\multicolumn{2}{c}{$Z_2^{U,D}$}\\
$h_u$, $h_d$ & nothing & nothing & nothing
\end{tabular}
\caption[breaking]{The sequential symmetry breaking pattern of $(S_3)^3$
symmetry.  The degeneracy between $\tilde{d}$ and $\tilde{s}$ is kept
until the
non-Abelian group $(Z_2^U \times S_3^D) /
Z_2$ is broken.  $Z_2^{U,D}$
symmetry remains unbroken until the last stage, which keeps up- and
down-quarks massless.}
\label{sequential}
\end{center}
\end{table}

The form of the Yukawa matrices presented above can be understood as a
consequence of the sequential breaking of the flavor symmetry, as shown in
Table~\ref{sequential}. Note that it is necessary to have different
flavons associated with each step in the sequence, since all the
components in a single irreducible multiplet typically become heavy at the
same time unless a fine-tuning is done.  For instance, we assume that one
chiral superfield transforming as $\phi({\bf \tilde{2}}, {\bf 1}_S,
{\bf 1}_S)$ acquires a vacuum expectation value (VEV) in its {\boldmath
$v$}$_1$ component, which breaks $S_3^Q$ to $Z_2^Q$ and
generates $h_t V_{cb}$.  A different chiral superfield transforming in the
same way, remains light at this stage, but acquires a VEV in its {\boldmath
$v$}$_2$ component at lower scale to break $Z_2^Q$ and generate
$h_t V_{ub}$.  Since $S_3^Q$ is  completely broken at this stage, a
single chiral superfield $\phi({\bf\tilde{2}}, {\bf 1}_A, {\bf 2})$ is split
into two $({\bf 1}_A, {\bf 2})$s under the remaining $(S_3^U\times
S_3^D)/Z_2$ symmetry; both acquire VEVs in their
{\boldmath $v$}$_1$ components to break this symmetry down to
$Z_2^{U,D}$.  Since $h_s$ and $h_s V_{cd}$ are generated at the same
stage of the symmetry breaking, there is a natural reason why the Cabibbo
angle is rather large.  The final stage of breaking is done by
another $\phi({\bf \tilde{2}}, {\bf 1}_A, {\bf 2})$ to generate $h_d$
(and by another $\phi({\bf \tilde{2}},{\bf \tilde{2}},{\bf 1}_S)$ to
generate $h_u$ in the up sector). Therefore, the $2\times 2$ block
in $Y_D$ has the structure $Y_D = Y_1 + Y_2$, where
\begin{equation}
Y_1 = \left( \begin{array}{cc}
      0 & a h_s \lambda \\ 0 & h_s
\end{array} \right),
\hspace{2cm}
Y_2 = \left( \begin{array}{cc}
      h_d & 0 \\ 0 & 0
      \end{array} \right).
\label{eq:y}
\end{equation}
Note that $Y_1$ preserves $Z_2^{U,D}$ symmetry.
Of course the other elements in $Y_2$ can be non-vanishing, but are
expected to be of order $h_d$ or less, and are irrelevant for our
purposes.  Similarly, the $2\times 2$ block in $Y_U$ is given by
$Y_U=Y'_1+Y'_2$, where
\begin{equation}
Y'_1 = \left( \begin{array}{cc}
      0 & a' h_c \lambda \\ 0 & h_c
\end{array} \right),
\hspace{2cm}
Y'_2 = \left( \begin{array}{cc}
      h_u & 0 \\ 0 & 0
      \end{array} \right).
\label{eq:y'}
\end{equation}
Note that $a$ and $a'$ are order one constants, with $a-a'=1$.

\section{Incorporating Lepton Sector}

The Higgs fields in the $(S_3)^3$ model transform nontrivially under
the flavor symmetry group.  Since the lepton fields acquire their
masses in the MSSM from the same Higgs fields as the quarks, the leptons
should transform under the same $(S_3)^3$ flavor symmetry.  We are led by
three principles in determining the precise transformation properties of
the lepton fields:
\begin{enumerate}
\item We do not allow any new flavor symmetries (e.g. new $S_3$ factors)
that arise only in the lepton sector. The only flavor symmetry in
the theory is $S_3^Q \times S_3^U \times S_3^D$.
\item We assign the transformation properties of the lepton fields so
that the charged lepton Yukawa matrix is similar to that of the
down quarks. This choice is suggested by the phenomenology.
\item We require that the most dangerous dimension-five operator
that contributes to proton decay, $(QQ)(QL)$, is forbidden in the $(S_3)^3$
symmetry limit.
\end{enumerate}
As we will see below, these principles are sufficient to completely
determine the transformation properties of the lepton fields.

Let us first consider the consequences of the first two conditions.
The down-quark Yukawa matrix is a coupling between the left-handed quark
fields $Q \sim ({\bf 1}_A+{\bf 2}, {\bf 1}_S, {\bf 1}_S)$ and the right-handed
down quark fields $D \sim ({\bf 1}_S, {\bf 1}_S, {\bf 1}_A + {\bf 2})$.
We know that the Yukawa matrix of the charged leptons is quite similar
to that of the down quarks, up to factors of order
three \cite{GJ} at high scales:
\begin{equation}
m_b \simeq m_\tau, \hspace{1cm} m_s \simeq \frac{1}{3} m_\mu,
\hspace{1cm} m_d \simeq 3 m_e.
\label{eq:3factors}
\end{equation}
Therefore, we look for an assignment of lepton transformation properties
that leads automatically to this observed similarity.  There are only
two possibilities:
\begin{equation}
\begin{array}{c|ccc}
& S_3^Q & S_3^U & S_3^D\\ \hline L & {\bf 1}_A+{\bf 2} & {\bf 1}_S &
{\bf 1}_S \\ E & {\bf 1}_S & {\bf 1}_S & {\bf 1}_A + {\bf 2}
\end{array}
\hspace{1cm}
\mbox{or}
\hspace{1cm}
\begin{array}{c|ccc}
& S_3^Q & S_3^U & S_3^D\\ \hline L & {\bf 1}_S & {\bf 1}_S & {\bf
1}_A+{\bf 2} \\ E & {\bf 1}_A + {\bf 2} & {\bf 1}_S & {\bf 1}_S
\end{array}
\ \ \ . \label{true}
\end{equation}
The third condition above allows us to distinguish between these
two alternatives. In the first assignment, the operator
$(Q_i Q_i) (Q_j L_j)$ is allowed by the $(S_3)^3$ symmetry, and
we have proton decay at an unacceptable rate.  Therefore, only the second
assignment in Eq.~(\ref{true}) satisfies all three criteria listed
above.  We could have obtained the same conclusion by considering
the $UUDE$ operators as well.

The remaining question that we need to answer is how the factors of
three in (\ref{eq:3factors}) enter in the Yukawa matrices.  One
plausible explanation is that they originate from fluctuations
in the order one coefficients that multiply the $(S_3)^3$ breaking
parameters which generate the quark and lepton Yukawa matrices.
As discussed in the previous section, it is quite likely that the
$2\times 2$ block in $Y_d$ is generated by two $\phi({\bf
\tilde{2}}, {\bf 1}_A, {\bf 2})$ breaking parameters.
An acceptable lepton Yukawa matrix is obtained by allowing
coefficients in both breaking parameters to deviate from unity by a factor of
three. Throughout this paper we take $Y_l = 3 Y_1 + \frac{1}{3} Y_2$.

\section{Uniqueness}\label{sec:unique}

In the basic (2,2) model introduced in the previous section, the quark
Yukawa matrices were taken as the only sources of flavor symmetry breaking.
Thus, the Higgs fields that spontaneously break the flavor symmetry group in
the corresponding high energy theory would come in exactly four
representations of $(S_3)^3$: $({\bf 2},{\bf 1}_A,{\bf 2})$,
$({\bf 1}_S,{\bf 1}_A,{\bf 1}_A)$, $({\bf 2},{\bf 2},{\bf 1}_S)$,
and $({\bf 2},{\bf 1}_S,{\bf 1}_S)$.  These correspond to the various blocks
of the quark Yukawa matrices.  We have assumed that no other representations
are involved in $(S_3)^3$ breaking in the full theory.

The issue that remains to be addressed is whether this picture
of the high energy theory is overly restrictive.   There are
$3^3-1 = 26$ nontrivial representations of $(S_3)^3$ that we could
have used had we tried to build an adequate high-energy theory of
flavor symmetry breaking at the start.  There is no reason to
assume {\em a priori} that a model cannot be constructed with
a different, more fundamental set of symmetry breaking parameters.
What we will show in this section is that there are in fact no simple
models of $(S_3)^3$ breaking involving symmetry breaking parameters that
are more fundamental than the ones adopted in the basic (2,2) model.
We will systematically exclude all the reasonable alternatives.  While the
basic (2,2) model served as an existence proof for a successful
$(S_3)^3$ model in Ref.~\cite{hallmur}, we will show here that the choice
of symmetry breaking parameters in this model is in fact unique.

We first would like to consider the class of models in which
there are no ``$(2,2)$'' representations, i.e., there are no fields
that transform as doublets under more than one $S_3$ group
at a time.  As a starting point, let us consider a simple toy
model that illustrates the phenomenological problems common to
models of this type.  The flavon content is
\beq
H_Q^{(i)}\sim({\bf 2},{\bf 1}_S,{\bf 1}_S) \,\,\,\,\,\,\,
H_U\sim({\bf 1}_S,{\bf 2},{\bf 1}_S) \,\,\,\,\,\,\,
H_D\sim({\bf 1}_S,{\bf 1}_S,{\bf 2}) \,\,\,\,\,\,\,
\eeq
where $i=1 \ldots 2$ labels two distinct doublets, and
\beq
\chi_1 \sim ({\bf 1}_S,{\bf 1}_A,{\bf 1}_S) \,\,\,\,\,
\chi_2 \sim ({\bf 1}_S,{\bf 1}_A,{\bf 1}_A) \,\,\,\,\,
\chi_3 \sim ({\bf 1}_A,{\bf 1}_A,{\bf 1}_S) \,\,\,\,\,
\chi_4 \sim ({\bf 1}_A,{\bf 1}_A,{\bf 1}_A)
\label{eq:start}
\eeq
The doublet $H$ fields were chosen to have the simplest
quantum number assignments possible.  The quantum number assignments
of the $\chi$ fields were chosen for a variety of reasons, that
will become clear in context below.   When the flavon fields acquire vevs,
the various blocks of the quark mass matrices are generated from
higher dimension operators.  The way in which this model fails is
instructive and will simplify the discussion of the other models to
follow, so we will proceed in some detail.

The two-by-two quark Yukawa matrices in this model are obtained
by taking the products $\sigma_2 H_Q H_U^T \sigma_2^T$ and
$\sigma_2 H_Q H_D \chi_U$, for the up and down sectors respectively.
Given this construction, we require two $H_Q$ fields in order to assure
a nonvanishing Cabibbo angle.  Let us denote the ratio of the vevs of
the $H$ and $\chi$ fields to an appropriate cutoff scale by $\epsilon$
and $\delta$, respectively.  If we take the $H_Q$ and $H_D$ to be of
the form
\beq
H_Q^{(i)} = \epsilon_Q \left[ \begin{array}{c} 1 \\ \lambda
                        \end{array} \right]
\,\,\,\,\,\,\,\,
H_D = \epsilon_D \left[ \begin{array}{c} \lambda \\ 1
                        \end{array} \right]
\eeq
then we obtain the two-by-two down Yukawa matrix
\beq
\epsilon_Q \epsilon_D \delta_1
\left[\begin{array}{cc} \lambda^2 & \lambda \\ \lambda & 1 \end{array}
\right]
\label{dtbt}
\eeq
where $\lambda \approx 0.22$ is the Cabibbo angle.  By setting
the combination $\epsilon_Q \epsilon_D \delta_1 \sim \lambda^5$, we
obtain the correct strange quark Yukawa coupling, assuming
$\tan\beta\sim 1$.

The down quark Yukawa coupling has not yet been generated, however, because
the matrix above has a zero eigenvalue.  This is a generic feature
of all models in which the down-strange Yukawa matrix is formed by taking
the product of two doublets.  Of course, the same problem arises in the
up-charm Yukawa matrix as well.  The way in which the up
and down Yukawa couplings are generated is through other operators that
are of higher order in the symmetry breaking.  Notice that we can obtain
a correction to the down-strange Yukawa matrix via the operator
$\chi_4 H_Q H_D^T \sigma_2^T$ which is of the form
\beq
\epsilon_Q \epsilon_D \delta_4
\left[\begin{array}{cc} 1 & \lambda \\ \lambda & \lambda^2 \end{array}
\right]
\eeq
If we take $\delta_4/\delta_2 \sim \lambda^2$, then we obtain a
correction to the (1,1) component of (\ref{dtbt}).  This is sufficient to
lift the zero eigenvalue at order
$\epsilon_Q \epsilon_D \delta_4 \sim \lambda^7$, which is
of the correct order in $\lambda$ to generate the down Yukawa coupling.
A similar mechanism occurs in the up sector as well.
If we take
\beq
H_U^{(i)} = \epsilon_Q \left[ \begin{array}{c} 1 \\ \lambda
                        \end{array} \right]
\eeq
then the up-charm Yukawa matrix $\sigma_2 H_Q H_U^T \sigma_2^T$
is given by
\[
\epsilon_Q \epsilon_U
\left[\begin{array}{cc} \lambda^2 & \lambda \\ \lambda & 1 \end{array}
\right]
\]
We take $\epsilon_Q \epsilon_U \sim \lambda^4$ in order to generate
the correct charm quark Yukawa coupling.  The zero eigenvalue in
this matrix is lifted via the operator $\chi_3 H_Q H_U^T$,
which is of the form
\[
\epsilon_Q \epsilon_U \delta_3
\left[\begin{array}{cc} 1 & \lambda \\ \lambda & \lambda^2 \end{array}
\right]
\]
If we take $\delta_3 \sim \lambda^4$, then this corrects the (1,1) entry
of the first operator, lifting the zero eigenvalue at order
$\lambda^8$, as desired.

The inescapable problems with the model above arise from
flavor changing neutral currents and proton decay considerations.  In
the first case, the off-diagonal elements in the squark mass matrices can
be constructed at first order in the symmetry breaking $\epsilon$ parameters
via the `$2^3$ invariant'
\beq
(\tilde{Q}^{*T} \sigma_3 \tilde{Q}) H_1
- (\tilde{Q}^{*T} \sigma_1 \tilde{Q}) H_2
\eeq
Here $\tilde{Q}$ is the squark doublet that transforms under the
same $S_3$ group as $H$, and we use subscripts to
denote the components of the $H$ doublet.   The resulting off-diagonal
elements are constrained by flavor changing neutral currents
processes, such that $\langle H_{Q \,2} \rangle < 0.05$,
$\langle H_{D \,2} \rangle < 0.05$, and
$\sqrt{\langle H_{Q \,2}\rangle\langle H_{D \,2}\rangle}<0.006$.
Given the form of the $H$ fields described above, this implies
that $\sqrt{\epsilon_Q \epsilon_D \lambda}<0.006$.  However, we saw
earlier that the strange quark Yukawa coupling
$h_s \sim \epsilon_Q \epsilon_D \delta_1$ is of order $\lambda^5$.
Hence we require $\delta_1 \approx 3$, which contradicts our assumption
that all the $\epsilon$'s and $\delta$'s are small symmetry breaking
parameters. In other words, for $\delta_1<1$, we cannot generate a large
enough strange quark Yukawa coupling if we are to simultaneously satisfy
the FCNC constraints.

The second problem is that there are operators in this model
that contribute to proton decay at an unacceptable level.
We can construct the representation $({\bf 2},{\bf 1}_S,{\bf 2})$ at order
$\epsilon_Q \epsilon_D$, which contributes to proton
decay via the operator $\frac{1}{M}(QQQL)$ at the same order.
This forces us to take $\epsilon_Q \epsilon_D < \lambda^7$, which
again makes it impossible to generate a large enough strange quark
Yukawa coupling.

What we have found is that it is impossible to generate large
enough down quark Yukawa couplings when the $H$ fields have
the simple transformation properties described above.  One way to remedy
this problem is to allow these fields to transform under more than
one $S_3$ group, so that the contributions to the off-diagonal elements
of the corresponding squark mass matrices occur at higher order,
and dangerous proton decay operators are sufficiently suppressed.  The
minimal modification of the model above with this property is a model in
which some or all of the $H$ fields transform as ${\bf 1}_A$'s
under an additional (or both remaining) $S_3$ groups.  Since
it is possible in this case to build the down-strange Yukawa matrix from
the product of $H_Q$ and $H_D$ alone, we will restrict our
discussion to models in which the both the down-strange and up-charm
Yukawa matrices are generated at second order in the symmetry-breaking
$\epsilon$ parameters.   Any model in which these matrices
are generated at higher order in the symmetry breaking will involve the
representations described below as composite operators, and at the
very least will be subject to the same constraints.  There are a finite number
of possibilities for the type of model of interest, and we will now outline
why each fails:

{\em case 1:} $H_D \sim ({\bf 1}_A,{\bf 1}_S,{\bf 2})$.  This
representation contributes to proton decay through the
operator $\frac{1}{M}(QQQL)$ at order $\lambda^2 \epsilon_D$.  Assuming
that none of the $\epsilon$'s are larger than order $\lambda$,
then $\epsilon_D \geq \lambda^4$ so that we can generate a large enough
strange quark Yukawa coupling.  Hence, the coefficient of the proton decay
operator is order $\lambda^6$ or greater.  This leads to an enhancement in
the proton decay rate by four orders of magnitude over that of the
basic (2,2) model, and this possibility is excluded.

{\em case 2:} $H_D \sim ({\bf 1}_S,{\bf 1}_A,{\bf 2})$.  To construct the
upper-left two by two block of the down quark mass matrix using this
representation we must also have
$H^{(i)}_Q \sim ({\bf 2},{\bf 1}_S,{\bf 1}_S)$.  (The alternative
choice $H_Q^{(i)} \sim ({\bf 2},{\bf 1}_S,{\bf 1}_A)$, for any $i$, is
excluded by the proton decay constraints, as we will discuss later.)  Notice
that we generate the (1,3) and (2,3) entries of the down quark matrix from the
product of $H_Q$ and $h_b$.  Thus we require $\epsilon_Q$ to be
order $\lambda^2$ or smaller if $V_{ub}$ and $V_{cb}$ are not to be
unacceptably large.  This forces us to take $\epsilon_D$ of
order $\lambda^3$ or larger if we are to generate an adequate strange
quark Yukawa coupling. Now the problem arises because $H_D$ also
contributes to the (3,1) and (3,2) entries of the down quark Yukawa matrix
at order $\epsilon_D$, which we have just argued is of order $\lambda^3$
or larger.   This gives us at least an order $\lambda$ rotation on the
right-handed down quark fields between the first and third generations.
In the basis where the quark Yukawa matrices are diagonal, this yields
a (3,1) entry in the right-handed down squark mass matrix of
order $\lambda M_1$ (Recall that the diagonal components of the squark
mass matrices $M_1$ and $M_3$ are unconstrained.)  This is in marginal
disagreement with the the bound ${(m^2)}^D_{13}/M_1^2< 0.1$ from
the $B^0$-$\overline{B}^0$ constraint.  A more serious problem arises
because there is also an order 1 rotation between the right-handed $b$
and $s$ fields, which spoils the degeneracy between the second and
first generation squarks;  since there is an order $\lambda$ rotation on
the right-handed squarks of the first two generations in this model, we
obtain a $(\delta^d_{RR})_{12}$ of order $\lambda$, which exceeds the bound
$(\delta^d_{RR})_{12} < 0.05$.

{\em case 3:} $H_D \sim ({\bf 1}_A,{\bf 1}_A,{\bf 2})$  In models that
include this representation we must also
have  $H_Q^{(i)} \sim ({\bf 2},{\bf 1}_S,{\bf 1}_S)$. (Again,
the alternative $H_Q^{(i)} \sim ({\bf 2},{\bf 1}_S,{\bf 1}_A)$, for any $i$,
is excluded by proton decay constraints, as discussed below.)   $H_Q$
contributes to the (1,3) and (2,3) entries of both the up and down Yukawa
matrices at order $\epsilon_Q$ and $\epsilon_Q h_b$ respectively, but
always with the largest component in the (3,1) entry.  Thus, this model
predicts $V_{ub}>V_{cb}$, and is therefore excluded.

Now that we have established that $H_D \sim ({\bf 1}_S,{\bf 1}_S,{\bf 2})$
is the only viable representation for $H_D$ (that has no more than 2 degrees of
freedom) we are forced to modify the $H_Q$ if we are going to evade the
problems of the simple model presented at the beginning of this
section.

{\em case 4:} $H_Q \sim ({\bf 2},{\bf 1}_A,{\bf 1}_A)$.  This representation
contributes at order $\epsilon_Q$ to the (1,3) and (2,3) entries of the down
quark Yukawa matrix.  Thus, $\epsilon_Q$ can be no larger than order
$\lambda^5$ if $V_{cb}$ and $V_{ub}$ are to be small enough.
In this case, we cannot generate a large enough strange quark Yukawa
coupling.

{\em case 5:} $H_Q \sim ({\bf 2},{\bf 1}_S,{\bf 1}_A)$.   This contributes to
proton decay via the $\frac{1}{M}(QQQL)$ operator at order
$\epsilon_Q$ and is immediately excluded.

{\em case 6:} $H_Q \sim ({\bf 2},{\bf 1}_A,{\bf 1}_S)$. We can construct the
spurion $({\bf 2}, {\bf 1}_S, {\bf 1}_A)$ at
order  $h_b \epsilon_Q \sim \lambda^3 \epsilon_Q$, which contributes to
proton decay via $\frac{1}{M}(QQQL)$ at the same order.
This forces us to take $\epsilon_Q \sim \lambda^4$ or smaller, which
in turn tells us that $\epsilon_D \sim \lambda$, in order to generate
a large enough strange quark Yukawa coupling.  However, this is
excluded by the flavor changing neutral current constraints for
$H_D \sim ({\bf 1}_S,{\bf 1}_S,{\bf 2})$.

The discussion above forces us to consider representations for
the $H$'s that have four degrees of freedom.  Let us consider
the following possibilities:

{\em case 7}: $H_D \sim ({\bf 2}, {\bf 1}_S, {\bf 2})$ .  We could imagine
constructing a model in which the down two by two yukawa matrix is a product
of $({\bf 2}, {\bf 1}_S, {\bf 2})$ and $h_b$.  However, $H_D$ now gives us a
contribution to proton decay at order $\epsilon_D$, and hence this
alternative is excluded.

{\em case 8}: Other Models without a $({\bf 2}, {\bf 1}_A, {\bf 2})$.
We still might hope to construct models without
a $({\bf 2}, {\bf 1}_A {\bf 2})$ if we can generate the upper two by two
block of the down quark Yukawa matrix at second order in the symmetry
breaking.  In such models, we require that the upper two by two block of the
up quark Yukawa matrix be generated at second order as well.  This
restricts us to models with the
representations (a) $({\bf 1}_A,{\bf 2},{\bf 2})$ and
$({\bf 2},{\bf 2},{\bf 1}_A)$
or (b) $({\bf 1}_S,{\bf 2},{\bf 2})$ and $({\bf 2},{\bf 2},{\bf 1}_A)$.  In
both cases, the down two by two block is generated by taking the product of the
two spurions, while the up matrix is generated by taking the product of
$({\bf 2},{\bf 2},{\bf 1}_A)$ and $h_b$.  Of course, we must also introduce a
$({\bf 2},{\bf 1}_S,{\bf 1}_S)$ if we are to generate sufficient $V_{ub}$ and
$V_{cb}$.  The problem with these models is that we can
also construct a $({\bf 2},{\bf 1}_S,{\bf 2})$ by taking a product of the same
spurions that generate the strange quark Yukawa coupling.  Hence,
there is a contribution to proton decay at order $\lambda^5$, and
these models are excluded.

We have seen above that we must have a $({\bf 2}, {\bf 1}_A, {\bf 2})$ as a
fundamental field in the model, and that the down and
strange Yukawa couplings are generated at first order
in the symmetry breaking.  If we require that the up quark
two by two block also be generated at first order (so that
the symmetry breaking is of a comparable magnitude) then
we must introduce a fundamental $({\bf 2},{\bf 2},{\bf 1}_s)$.  Since we
cannot generate large enough values for $V_{ub}$ and $V_{cb}$ by taking
products of these representations alone, we must also introduce
a $({\bf 2},{\bf 1}_s,{\bf 1}_s)$.  Finally, we generate $h_b$ most
economically with a $({\bf 1}_S,{\bf 1}_A,{\bf 1}_A)$.  Thus, we have
arrived uniquely at the basic (2,2) model described in sections 3 and 4.

\section{Phenomenology}

In the previous sections, we showed that the transformation
properties of both the lepton and flavon fields are uniquely
determined in the $(S_3)^3$ model, if all possible nonrenormalizable
operators are generated at the Planck scale.  With these results
at hand, we may now consider lepton flavor violation and proton
decay, two topics that depend crucially on the extension of the
model to the lepton sector.   We will also consider
the bounds from CP-violating processes, which were
not covered in the original paper.  We demonstrate that the
model is indeed viable phenomenologically.  (Whenever we discuss
numerical estimates, we take a representative
choice $\tan \beta \simeq 2$.)  In addition,
we show that the model predicts the dominance of the $K e$ proton
decay mode over the $K\mu$ or $\pi e$ mode, which is in sharp contrast
to the situation in supersymmetric or non-supersymmetric GUTs.

\subsection{Lepton Flavor Violation}

The strongest constraint on lepton flavor violation comes from the
non-observation of the $\mu \rightarrow e \gamma$ decay mode.  In our
model, the contribution of the off-diagonal term in the purely
left-handed slepton mass matrix (the LL matrix) is small
enough ($\sim h_s^2 \lambda \sim 1 \times 10^{-7}$) to avoid the experimental
constraint for any value of $m_{\tilde{l}}$ above the LEP bound.  The
stringent limits come from the purely right-handed slepton mass matrix (RR)
and the left-right (LR) matrix, which we discuss in this section. We follow
the notation of Barbieri, Hall and Strumia \cite{BHS} below.  For simplicity,
we work in the approximation where the exchanged neutralino is a pure bino
state.

The one-loop slepton and bino exchange diagram that picks up the off-diagonal
(2,1) component in LR mass matrix generates the operator
\begin{equation}
{\cal O} = \frac{e}{2} F_2(m^2_R, m^2_L, M_1^2)
      \,\,\, \bar{e}_R i\sigma^{\mu\nu} \mu_L
      F_{\mu\nu} \, ,
\label{eq:mego}
\end{equation}
where $F_{\mu\nu}$ is the electromagnetic field strength and
\begin{equation}
F_2 (m^2_R, m^2_L, M_1^2) =
      \frac{\alpha_Y}{4\pi} (m^2_{LR})_{21}
      \frac{ G_2 (m_R^2, M_1^2) - G_2 (m_L^2, M_1^2)}{m_R^2 - m_L^2},
\end{equation}
with
\begin{equation}
G_2 (m^2, M^2) = \frac{M}{2(m^2 - M^2)^3}
      \left[ m^4 - M^4 - 2 m^2 M^2 \ln \left(\frac{m^2}{M^2}\right)
            \right].
\end{equation}
The decay width is given by
\beq
\Gamma( \mu \rightarrow e \gamma) = \frac{\alpha}{4} m_\mu^3 |F_2|^2 \,,
\eeq
and the bound Br$( \mu \rightarrow e \gamma) < 4.9 \times 10^{-11}$ implies
$|F_2| < 2.6 \times 10^{-12}$~GeV$^{-1}$.  In order to compare this bound to
the prediction of our model, let us take $m_R = m_L = m = 300$~GeV and
$M_1 = 100$~GeV as a representative case. Note that $F_2$ is at its maximum
around $M_1 \sim m/2$, and our choice gives almost the largest possible $F_2$
for fixed $m$.  We obtain
\begin{equation}
\frac{(m^2_{LR})_{21}}{m^2} < 1.0 \times 10^{-5}
\label{eq:cons}
\end{equation}
for this choice of parameters.  In our model, the (2,2) and (1,2)
elements in $Y_l$ belong to the same irreducible multiplet, and
diagonalization of $Y_l$ also diagonalizes LR mass matrix at $O(h_s
\lambda)$.  The term which may not be simultaneously diagonalizable
comes from the piece $Y_2 \sim h_d$, and hence
\begin{equation}
(m^2_{LR})_{21} \sim m_d \lambda A,
\end{equation}
where $m_d$ is the down quark mass evaluated at the Planck scale $m_d
\simeq 10~\mbox{MeV}/ 3$, and $A$ is a typical trilinear coupling. If
we take $A \sim 100$~GeV, then $(m^2_{LR})_{21}/m^2 \sim 0.8 \times
10^{-6}$ and the constraint (\ref{eq:cons}) is easily satisfied.  Since
we have consistently allowed an order 3 ambiguity in estimating the
coefficients of various operators, and we do not know the precise
value of $m_d$, the actual constraint is weaker than the one
considered above.  The (1,2) element in the LR mass matrix contributes
in exactly the same way as the (2,1) entry, except that the chiralities
of the electron and muon in eq.~(\ref{eq:mego}) are flipped.  Hence,
the (1,2) element is subject to the same constraint, which again is
clearly satisfied in our model.

The RR mass matrix also contributes to the operator
in (\ref{eq:mego}).  In this case, the function
$F_2$ is given by
\begin{equation}
F_2 = \frac{\alpha_Y}{4\pi} m_\mu (m^2_{RR})_{12}
      \frac{\partial G_1(m^2, M_1^2)}{\partial m^2},
\end{equation}
where
\begin{eqnarray}
\lefteqn{
\frac{\partial G_1(m^2, M_1^2)}{\partial m^2} =
\frac{1}{6\,( {m^2} - {M^2} )^5} \times
      } & \nonumber \\
& \left( {m^6} - 9\,{m^4}\,{M^2} - 9\,{m^2}\,{M^4} +
       17\,{M^6} + (18\,{m^2}\,{M^4}+
       6\,{M^6})\,\log ({{{m^2}}\over {{M^2}}}) \right) .
\end{eqnarray}
This is a monotonically decreasing function of $M$ for fixed $m$.
For the bino and slepton masses chosen earlier, we
obtain the bound
\begin{equation}
\frac{(m^2_{RR})_{12}}{m^2} < 0.023,
\label{eq:mrrc}
\end{equation}
while in our model
\begin{equation}
(m^2_{RR})_{12}/m^2 \simeq h_t V_{cb} \lambda \sim 0.009  \,\,.
\end{equation}
The bound on the (1,2) element of the RR matrix is easily satisfied in
our model.  Note had we chosen the option of generating
$V_{cb}$ and $V_{ub}$ in the down sector, as discussed in section~3,
we would have obtained a much smaller off-diagonal element $(m^2_{RR})_{12}$.
In this case the breaking parameter transforms as
$\phi({\bf \tilde{2}}, {\bf 1}_A, {\bf 1}_A)$ rather
than $\phi({\bf \tilde{2}}, {\bf 1}_S, {\bf 1}_S)$, and does not
contribute to the RR mass matrix at the first order in $\phi$.  The leading
contribution is then
\begin{equation}
(m^2_{RR})_{12}/m^2 \simeq h_b^2 V_{cb} \lambda \sim 3 \times 10^{-6} \,\,,
\end{equation}
and the constraint (\ref{eq:mrrc}) is again easily satisfied.

Finally, there is another contribution to $\mu\rightarrow e \gamma$ from the
mixing to the third generation sleptons.  Since the third generation scalars
can be non-degenerate with the first and second generation ones, GIM
cancellation does not occur. This is similar to the situation in the minimal
SO(10) model \cite{DH,BHS}, where the third generation sleptons are much
lighter than the others, while the rotations to the quark mass eigenstate
basis are CKM-like.  Recall that in the minimal SO(10) model, the flavor
changing factors in the amplitudes are $V_{td} V_{ts}$, whereas in our model
they are $V_{ub}V_{cb}$, which is about a factor of three smaller.

To summarize, the $(S_3)^3$ model is clearly consistent with
the experimental bounds on lepton flavor violation, and predicts $\mu
\rightarrow e\gamma$ at a rate just beyond the current limit.

\subsection{CP Violation}

If we allow arbitrary phases in the symmetry breaking parameters, the
squark mass matrices can have complex elements which contribute to
CP violating effects.  The most stringent limit on these phases comes
from $\epsilon'$, and has been studied by Gabrielli, Masiero and Silvestrini
\cite{GMS}.  If we consider the most extreme case imaginable, where all
the off-diagonal elements of the down squark mass matrices in our model
are purely imaginary, then we find
\begin{eqnarray}
&&\mbox{Im}\left( \frac{(m_{LR}^2)_{12}}{m^2} \right) \simeq h_d \lambda
      \simeq 9 \times 10^{-6}, \\
&&\mbox{Im}\left( \frac{(m_{LL}^2)_{12}}{m^2} \right) \simeq h_t V_{cb} \lambda
      \simeq 0.009, \\
&&\mbox{Im}\left( \frac{(m_{RR}^2)_{12}}{m^2} \right) \simeq h_s^2 \lambda
      \simeq 1 \times 10^{-7} \,\, .
\end{eqnarray}
in the basis where the quark Yukawa matrices are diagonal.
The constraints on these elements are
\begin{eqnarray}
&&\left|\mbox{Im}\left( \frac{(m_{LR}^2)_{12}}{m^2} \right)\right|
      < 2\times 10^{-5},
      \\
&&\sqrt{\mbox{Im}\left( \frac{(m_{LL,RR}^2)_{12}}{m^2} \right)^2}
      < 3\times 10^{-3},
      \\
&&\sqrt{\mbox{Im}\left( \frac{(m_{RR}^2)_{12}}{m^2}
      \frac{(m_{LL}^2)_{12}}{m^2} \right) }
      < 2\times 10^{-4} ,
\end{eqnarray}
for $m_{\tilde{q}} \simeq m_{\tilde{g}} \simeq 500$~GeV.  One can see
that all constraints are easily satisfied.  In addition, there are a
number of factors that make the actual bounds on our model weaker:
(1) This analysis is valid only up to the unknown factors of order one that
multiply the symmetry breaking operators.  (2) The renormalization group
running of soft SUSY breaking masses always tends to make the diagonal
elements in the LL, RR, and LR matrices larger at lower energies, so that
the true constraints on our model are generally weaker than those given
above. (3) The elements $V_{cb}$ and $V_{ub}$ could instead be generated in
the down sector, in which case
$(m_{LL}^2)_{12}/m^2 \simeq h_b^2 V_{cb} \lambda $,
and the SUSY contribution to $\epsilon '$ becomes negligible.
(4) We have no reason to expect that all the off-diagonal elements of the
squark mass matrices in our model are purely imaginary, as we have assumed to
obtain these bounds.

As in the case of $\mu \rightarrow e\gamma$, there is also a contribution
to $\epsilon'$ from the third generation squarks, similar to the
minimal SO(10) case \cite{DH}.  This falls within an acceptable range.
Hence, it is reasonable to conclude that the model is consistent with
the observed CP violating phenomenology.

\subsection{Proton Decay}

Since we have assumed throughout this paper that all possible
nonrenormalizable operators are generated at the Planck scale,
the task of studying proton decay in our model is a simple one.
We first write down all the possible dimension-five operators
that contribute to proton decay and identify their transformation
properties under $(S_3)^3$.  The coefficients can be estimated as the
product of Yukawa couplings that will produce the desired symmetry
breaking effect.  A list of possible operators and their
coefficients is given in Tables~\ref{pdtable1} and \ref{pdtable2}.
\begin{table}
\centerline{
\begin{tabular}{c|c|c}
operator & $(S_3)^3$ representation & largest coefficient\\ \hline
$(Q_3 Q_3)(Q_i L_3)$ & $({\bf 2}, {\bf 1}_S, {\bf 1}_A)$
      & $h_s h_d$ or $h_b h_c h_u$\\
$(Q_3 Q_3)(Q_i L_i)$ & $({\bf 2}, {\bf 1}_S, {\bf 2})$
      & $h_b h_s$\\
$(Q_i Q_i)(Q_3 L_i)$ & $({\bf 2}, {\bf 1}_S, {\bf 2})$
      & $h_b h_s$\\
$(Q_i Q_i)(Q_3 L_i)$ & $({\bf 1}_A, {\bf 1}_S, {\bf 2})$
      & $h_t h_b h_s \lambda V_{cb}$\\
$(Q_i Q_i)(Q_3 L_3)$ & $({\bf 2}, {\bf 1}_S, {\bf 1}_A)$
      & $h_s h_d$ or $h_b h_c h_u$\\
$(Q_i Q_i)(Q_3 L_3)$ & $({\bf 1}_A, {\bf 1}_S, {\bf 1}_A)$
      & $h_s h_d$\\
$(Q_i Q_i)(Q_i L_3)$ & $({\bf 2}, {\bf 1}_S, {\bf 1}_A)$
      & $h_s h_d$ or $h_b h_c h_u$\\
$(Q_i Q_i)(Q_i L_i)$ & $({\bf 2}, {\bf 1}_S, {\bf 2})$
      & $h_b h_s$
\end{tabular}
}
\caption{A complete list of possible baryon-number violating
dimension-five operators involving doublet fields.  All other operators
vanish because of symmetry reasons.  $Q_i$ and $L_i$
generically refer to either first- or second-generation fields, while
$Q_3$ and $L_3$ refer to third-generation fields.  For operators that
are multiplets under the $(S_3)^3$ symmetry, the coefficient of the
largest component is given.}
\label{pdtable1}
\end{table}

\begin{table}
\centerline{
\begin{tabular}{c|c|c}
operator & $(S_3)^3$ representation & largest coefficient\\ \hline
$t u_i b \tau$ & $({\bf 1}_A, {\bf 2}, {\bf 1}_A)$
      & $h_t h_b h_c V_{ub}$\\
$t u_i b e_i$ & $({\bf 2}, {\bf 2}, {\bf 1}_A)$
      & $h_b h_c$\\
$t u_i d_i \tau$ & $({\bf 1}_A, {\bf 2}, {\bf 2})$
      & $h_c h_s \lambda$\\
$t u_i d_i e_i$ & $({\bf 2}, {\bf 2}, {\bf 2})$
      & $h_c h_s$\\
$u c b \tau$ & $({\bf 1}_A, {\bf 1}_A, {\bf 1}_A)$
      & $h_t^2 h_b V_{cb} V_{ub}$\\
$u c b e_i$ & $({\bf 2}, {\bf 1}_A, {\bf 1}_A)$
      & $h_t h_b V_{cb}$\\
$u c d_i \tau$ & $({\bf 1}_A, {\bf 1}_A, {\bf 2})$
      & $h_t h_s V_{cb} \lambda$\\
$u c d_i e_i$ & $({\bf 2}, {\bf 1}_A, {\bf 2})$
      & $h_s$\\
\end{tabular}
}
\caption{A complete list of possible baryon-number violating
dimension-five operators involving singlet fields.  All other operators
vanish because of symmetry reasons. $u_i$, $d_i$, and
$e_i$ generically refer to either first- or second-generation fields, while
$t$, $b$ and $\tau$ refer to the third-generation fields.  The only
possible product of the form $u_i u_i$ is $u c$ because of the
anti-symmetry in the color indices.  For operators that are multiplets
under the $(S_3)^3$ symmetry, the coefficient of the largest component
is given.}
\label{pdtable2}
\end{table}

As we can see from the tables, the most important operator involving
left-handed fields is
$ (Q_i Q_i)(Q_i L_i)/ M_*$, where $M_* \equiv M_{Pl}/\sqrt{8\pi}$ is
the reduced Planck mass, and where parentheses indicate a contraction
of SU(2) indices.  This operator transforms as
a $({\bf 2}, {\bf 1}_S, {\bf 2})$ under $(S_3)^3$, and therefore has a
coefficient of order $h_b h_s$.    The two other operators that have
the same coefficient, $(Q_3 Q_3)(Q_i L_i)$ and  $(Q_i Q_i)(Q_3 L_i)$,
each involve third generation fields, and therefore contribute
to proton decay at a rate that is further suppressed by small CKM angles.
The coefficients for all four components of the leading operator are
given by
\begin{eqnarray}
O &=&  \frac{c}{2} \frac{h_b}{M_*}
      \left[ h_s (Q_2 Q_2) (Q_1 L_1)
            - h_s \lambda (Q_1 Q_1) (Q_2 L_1) \right. \nonumber \\
& & \left.
            - {\cal O}(h_d) (Q_2 Q_2)(Q_1 L_2)
            + h_d (Q_1 Q_1) (Q_2 L_2) \right] . \label{dimen5}
\end{eqnarray}
The overall coefficient $c$ is a number of ${\cal O}(1)$, and the factor
of $1/2$ has been included to compensate for the symmetry of each term
under the interchange of either the two $Q_1$'s or two $Q_2$'s.  The
striking feature of this multiplet is that the operators involving
first generation lepton fields $L_1$ have larger coefficients than
those involving second generation fields $L_2$.

The only operator involving right-handed fields that appears potentially
dangerous is $h_s (u c) (s e - \lambda s \mu)$.  However, the
contribution of this operator to proton decay is in fact negligible.
Since all fields are right-handed, their flavor cannot change
via $W$-exchange.  A $c$ squark can change to a $u$ squark via a flavor
off-diagonal element of the RR squark mass matrix, and the
resulting operator can be dressed by a gluino. Since the off-diagonal
entry is $(m^2_{RR})_{12}/m^2 \simeq h^2_c \lambda $, the effective
coefficient is smaller than $h_s h_c$.  Thus, this operator is
negligibly small compared to the leading left-handed operator discussed
above.  The same can be shown for all the remaining operators in
Table~\ref{pdtable2}.

What we have concluded based on the leading operator is that our model
favors proton decay to $\nu_e$ and $e$ over decay to $\nu_\mu$ and $\mu$.
This result is in striking contrast to the situation in grand unified
theories, and provides a counterexample to a common theoretical prejudice.
Usually it is assumed that the operator involving $L_1$ should be most
strongly suppressed because it only violates the electron's
U(1) chiral symmetry through the tiny electron Yukawa coupling.
Nevertheless, the operator involving the electron in our model
is larger than the operator involving the muon.

This unusual feature can easily be understood by considering the
residual $Z_2^{U,D}$ symmetry that is present
when the Yukawa couplings of first generation fields are set to
zero.  Suppose that the all the breaking parameters except
$h_u$ and $h_d$ are present.  This leaves a $Z_2^{U,D}$ symmetry, where
the fields have the charge assignments given in the
Table.~\ref{chas}. $Z_2^{U,D}$ is a diagonal subgroup of $Z_2^U$ and
$Z_2^D$.  What we learn from this table is that the first generation $L$
field is even under this $Z_2$, while the second and third generation
$L$ fields are odd.  Thus, in the symmetry limit, $Z_2^{U,D}$ forbids the
dimension-five operators containing $L_2$ but allows those involving $L_1$.
The same argument forbids operators involving $L_3$ in the $Z_2^{U,D}$
symmetry limit.

\begin{table}
{\scriptsize
\begin{tabular}{c|ccc|ccc|ccc|cc}
&$Q_1$, $E_1$ & $Q_2$, $E_2$ & $Q_3$, $E_3$& $U_1$ & $U_2$ & $U_3$
      & $D_1$, $L_1$ & $D_2$, $L_2$ & $D_3$, $L_3$
      & $H_U$ & $H_D$ \\ \hline
$Z_2^{U,D}$ & $+$ & $+$ & $+$ & $+$ & $-$ & $-$ & $+$ & $-$ & $-$
      & $-$ & $-$
\end{tabular}
}
\caption{Charge assignments of MSSM fields under the 
$Z_2^{U,D}$ symmetry.} \label{chas}
\end{table}

The predicted nucleon decay modes are obtained from Eq.~(\ref{dimen5})
by `dressing' the two-scalar-two-fermion operators with wino
exchange.\footnote{In principle, one can dress the operator
$uds\nu$ by gluino exchange.  Note that this operator is an SU(3) flavor
singlet because of the total anti-symmetry in the color indices.  The
gluino dressing can possibly lead to the four-fermion operator $(ud)(s\nu)$,
$(ds)(u\nu)$ or $(su)(d\nu)$.  However, all these operators have mixed
symmetry under flavor SU(3).  Because of the high degeneracy among the
squarks in our model, flavor SU(3) is a very good symmetry even with
the squarks.  Therefore, the four-fermion operators are suppressed by
the small non-degeneracy among the squarks, and hence are negligible.
There is a possibility that gluino dressing of operators involving
third-generation fields may be important.  But their coefficients are
already as small as the one we discuss here, and they need to pick up
smaller mixing angles and hence are negligible.} Below, we follow the
notation of Hisano, Murayama and
Yanagida \cite{HMY}.  The first operator $c h_b h_s (Q_2 Q_2) (Q_1
L_1)/M_*$ gives us the following four-fermion operators\footnote{In the
following discussion, we assume that the Cabibbo mixing originates from
the down sector, i.e. $a=1$, $a'=0$ in eq.~(\ref{eq:y}) and (\ref{eq:y'}).
However we checked that all the results remain the
same even when the Cabibbo mixing comes from both the down and the up
sectors, or even solely from the up sector.}
\begin{equation}
{\cal L} = \frac{\alpha_W}{2\pi} \frac{c h_b h_s}{M_*} \left[
      -\lambda (du) (s\nu_e) - \lambda (su) (d\nu_e) \right]
      ( f(c,e) + f(c,d) ).
\label{eq:firstop}
\end{equation}
where $f$ is the ``triangle diagram factor'' \cite{NA}, a function
of the wino and scalar masses:
\begin{equation}
f(i,j) = \frac{M_2}{m_i^2 - m_j^2}
      \left( \frac{m_i^2}{m_i^2 - M_2^2} \ln \frac{m_i^2}{M_2^2}
      - \frac{m_j^2}{m_j^2 - M_2^2} \ln \frac{m_j^2}{M_2^2} \right).
\end{equation}
and where parentheses now indicate the contraction of spinor indices.
In the limit where $M_2 \ll m$, this function is well approximated
by  $f \simeq M_2/m^2$. The second term in (\ref{dimen5}),
$- c h_b h_s \lambda (Q_1 Q_1) (Q_2 L_1)/M_*$, gives us
the four-fermion operators
\begin{equation}
{\cal L} = \frac{\alpha_W}{2\pi} \frac{c h_b h_s}{M_*} \left[
      \lambda (du) (s\nu_e) + \lambda (su) (ue) \right]
      ( f(c,e) + f(u, d) ).
\label{eq:secondop}
\end{equation}
All the fields in (\ref{eq:firstop}) and (\ref{eq:secondop}) are in the
mass eigenstate basis, and terms of higher orders in $\lambda$ have been
neglected.  The sum of these results is given by
\begin{equation}
{\cal L} = \frac{\alpha_W}{2\pi} \frac{c h_b h_s}{M_*} \left[
       - \lambda (su) (d\nu_e) + \lambda (su) (ue) \right]
       ( f(c,e) + f(c, d) ).
\label{eq:sumres}
\end{equation}
Here we have used the fact $f(c,d) = f(u,d)$ to good accuracy.  Note that
there is a precise cancellation between $(du)(s\nu_e)$ operators
in (\ref{eq:firstop}) and (\ref{eq:secondop}). This has an important
effect on the relative branching ratio between the charged lepton
and neutrino mode, as we will discuss below.

The ratios of decay widths can be estimated using the chiral Lagrangian
technique \cite{CWH,CD,HMY}.  We find
\begin{eqnarray}
\lefteqn{
\Gamma(p\rightarrow K^+ \bar{\nu}_e) : \Gamma(p\rightarrow K^0 e^+)
      : \Gamma(n \rightarrow K^0 \bar{\nu}_e) } \nonumber \\
& & = \left| \frac{2m_p}{3m_B}D\right|^2
      : \left|1 - \frac{m_p}{m_B}(D-F) \right|^2
      : \left|1 - \frac{m_n}{3m_B}(D-3F) \right|^2
= 0.4 : 1 : 2.7 . \nonumber \\
\label{eq:ratios}
\end{eqnarray}
where we have taken $m_B \equiv 1150$~MeV$\simeq m_\Sigma \simeq m_\Lambda$,
$D=0.81$ and $F=0.44$.  We have stressed earlier that the dominance of
the electron over the muon mode is a remarkable feature of this model,
which can never happen in SUSY-GUTs \cite{HMY,Aram,BB}.  In addition,
what is remarkable about the result in (\ref{eq:ratios}) is that
the proton's charged lepton decay mode dominates over the neutrino mode.
This is a consequence of the cancellation of the $(du)(s\nu_e)$ operator
in (\ref{eq:sumres}).  The dominance of $p\rightarrow K^0 e^+$ over
$p\rightarrow K^+\overline{\nu}_e$ is rarely the case in SUSY-GUTs.

Finally, we come to the overall rate. We find
\begin{eqnarray}
\lefteqn{
\tau(n \rightarrow K^0 \bar{\nu}_e) } \nonumber \\
& = & 4 \times 10^{31} \mbox{ years}
      \left| \frac{1}{c} \, \frac{0.003\mbox{ GeV}^3}{\xi} \,
		\frac{0.81}{A_S} \, \frac{5}{(1+\tan^2 \beta)} \,
      \frac{\mbox{TeV}^{-1}}{f(c,e)+f(c,d)} \right|^2 \, .
\label{eq:rate}
\end{eqnarray}
This result includes the effect of running the dimension-five operator
between the Planck scale and $m_Z$, (a factor of $A_S = 0.81$ in the
amplitude if $m_t = 175$~GeV, $\tan \beta = 2$, $\alpha_s (m_Z) = 0.12$)
and between $m_Z$ and $m_n$ (a factor of $0.22$ in the amplitude).  In
the expression above, $\xi$ is the hadronic matrix element of the
four-fermion operator evaluated between nucleon and kaon states; its exact
value is rather uncertain, but is estimated to be within the range
$\xi = 0.003$--0.03~GeV$^3$.  If we take that $M_2 \sim
100$~GeV and $m_{\tilde{q}} \sim 700$~GeV, and $m_{\tilde{l}} \sim 300$~GeV,
then the triangle functions $f(c,e)+f(c,d) \sim (1.8\mbox{ TeV})^{-1}$.
Thus, if $c=1$ and $\xi=0.003$ GeV$^3$, we obtain a mean lifetime
$12.7\times 10^{31}$ years, which can be compared to the experimental bound,
$\tau(n \rightarrow K^0 \bar{\nu}_e) > 8.6 \times 10^{31}$~years.
It is interesting to note that the coefficient $4 \times 10^{31}$
in (\ref{eq:rate}) would be the same in the minimal SU(5) SUSY-GUT
with an extremely large color-triplet Higgs mass $M_{H_C} = 10^{17}$~GeV.
Thus, the rate in our model is roughly comparable.  Overall,
the $(S_3)^3$ symmetry gives us just enough suppression of dimension-five
operators to evade the current bounds, so the model is phenomenologically
viable.  Since the SuperKamiokande experiment is expected to extend
Kamiokande's current reach by another factor or 30, there is a very good
chance that the $n\rightarrow K^0\overline{\nu}_e$  mode may be seen.  It
is an exciting prediction of this model that the $p \rightarrow K^0 e^+$
and $K^+ \bar{\nu}_e$ modes are likely to be seen at the same time
because their rates are close to each other, as we saw
in eq.~(\ref{eq:ratios}).

\section{Conclusions}

We have shown in this paper how to incorporate the leptons in
the $(S_3)^3$ model presented in Ref.~\cite{hallmur}.  By assuming
that all possible nonrenormalizable operators are generated at the
Planck scale, we showed that the transformation properties of both
the leptons and the flavor symmetry breaking fields could be uniquely
determined. We then demonstrated that the phenomenological constraints
from lepton flavor violation, CP violation, and proton decay are
indeed satisfied in our model.  The most striking prediction that emerged
from our analysis is the dominance of proton decay to final states
involving first generation lepton fields, unlike the case in SUSY GUTs.
We showed that the ratios of decay widths for the largest modes
$n \rightarrow K^0 \bar{\nu}_e$, $p\rightarrow K^+ \bar{\nu}_e$,
and $p\rightarrow K^0 e^+$  are approximately  0.4 :: 1 :: 2.7.   Given
our estimate of the total rate, we pointed out that all three modes may
be within the reach of the SuperKamiokande experiment and could well be
discovered simultaneously.

\begin{center}
{\bf Acknowledgments}
\end{center}

This work was supported in part by the Director, Office of
Energy Research, Office of High Energy and Nuclear Physics, Division of
High Energy Physics of the U.S. Department of Energy under Contract
DE-AC03-76SF00098 and in part by the National Science Foundation under
grant PHY-90-21139.






\end{document}